## ASTRONOMY

# Discovery of diffuse optical emission lines from the inner Galaxy: Evidence for LI(N)ER-like gas

D. Krishnarao[1]*, R. A. Benjamin[2], L. M. Haffner[3,4,1]

Optical emission lines are used to categorize galaxies into three groups according to their dominant central radiation source: active galactic nuclei, star formation, or low-ionization (nuclear) emission regions [LI(N)ERs] that may trace ionizing radiation from older stellar populations. Using the Wisconsin H-Alpha Mapper, we detect optical line emission in low-extinction windows within eight degrees of Galactic Center. The emission is associated with the 1.5-kiloparsec-radius "Tilted Disk" of neutral gas. We modify a model of this disk and find that the hydrogen gas observed is at least 48% ionized. The ratio [NII] λ6584 angstroms/Hα λ6563 angstroms increases from 0.3 to 2.5 with Galactocentric radius; [OIII] λ5007 angstroms and Hβ λ4861 angstroms are also sometimes detected. The line ratios for most Tilted Disk sightlines are characteristic of LI(N)ER galaxies.

## INTRODUCTION

Evidence for ionized gas in galaxies has existed since 1909 when optical emission lines were first detected in the spectra of a "spiral nebula" (1). Motivated by the discovery that the redshift of spectral lines of galaxies correlated with their distance (2), teams at Mt. Wilson Observatory and Lick Observatory began programs to obtain spectra for a large sample of these galaxies. Early on, both groups detected [OII] λ3727-Å emission from diffuse ionized gas in the inner regions ($R < 2$ kpc) of several galaxies (3, 4). Subsequent observations showed that this gas was characterized by a [NII] λ6584 Å/Hα λ6563 Å line ratio greater than unity, as opposed to the value of 0.3 seen in ionized gas surrounding HII regions in galaxy disks (5).

Emission line studies of ionized gas in subsequent decades focused on the nuclear regions ($R < 0.5$ kpc) of galaxies, leading to the classification of galaxies according to their emission line ratios. The line ratios [NII] λ6584 Å/Hα λ6563 Å and [OIII] λ5007 Å/Hβ λ4861 Å were found to be particularly useful in classifying galaxies according to the principal source of ionization (6). When combined with models of the radiation field, these line ratios allowed galaxies to be classified into one of three categories according to the dominant source of radiation: star formation, active galactic nuclei, or low-ionization (nuclear) emission line regions [LI(N)ERs] (7–9). This third category was introduced by Heckman (10); possible sources of ionization included shocks, cooling flows, and photoionization by evolved stellar populations (11).

Modern investigations of LI(N)ER emission (12–14) have resurrected interest in the radial extent of this gas, leading to the suggestion that the "nuclear" designation be dropped. Supporting this view, similar emission line ratios are found in the extended regions of elliptical and early-type galaxies, which also show evidence for an ultraviolet (UV) upturn. This UV upturn was first detected in the inner regions of M31 by Code (15), who speculated that it could be responsible for the previously observed diffuse ionized gas (16, 17). Stellar evolution and photoionization modeling of evolved stellar populations show that both the UV upturn and the optical emission line ratios may be explained by a population of hot old low-mass evolved stars [e.g., (18) and references therein]. There are several classes of potential ionizing sources, e.g., post–asymptotic giant branch (AGB) stars, AGB manqué stars, hot white dwarfs, pre-planetary nebula stars, and extreme horizontal branch/subdwarf OB stars, but the relative contributions of these sources to the total ionizing flux have not been definitively established. Since many of these sources are faint, it is not possible to resolve them in extragalactic systems. Even in M31—up until now, the nearest LI(N)ER—it is extremely challenging to detect individual sources and establish their contribution to the hydrogen-ionizing flux.

Here, we report the first detection of optical emission LI(N)ER-type gas in the inner part of the Milky Way Galaxy using the Wisconsin H-Alpha Mapper (WHAM) (19, 20). Our measurement of these optical emission lines in the inner Galaxy is possible due to two fortuitous circumstances. First, the neutral gas layer between Galactocentric radii of 0.5 kpc < $R_G$ < 1.5 kpc is tilted (21), extending more than 5° below the Galactic plane in Galactic longitudes $l = +10°$ to 0°. Second, this structure aligns with several low-extinction directions toward Galactic Center around $(l,b) = (1°$ to 5°, $−3°$ to $−6°)$, including Baade's Window. This tilted gas structure is distinct from the "Central Molecular Zone" (CMZ) interior to $R_G \sim 0.5$ kpc, which is also tilted, but with a different rotational axis.

We find three principal results. First, the optical line ratios for this gas are unlike anywhere else in the Milky Way Galaxy but typical of emission from LI(N)ER systems. Second, the inferred mass of ionized gas is much higher than predicted by current hydrodynamical models of gas flowing in a Milky Way–barred potential. A tilted geometrical model of the neutral gas modified to include an ionized component suggests that the atomic (nonmolecular) gas in the inner Galaxy is more than 50% ionized. And third, the [NII] λ6584 Å/Hα λ6563 Å ratio increases with Galactocentric radius, suggesting a radiation field that changes with position in the inner Galaxy. In the future, these observations may be compared with spatially resolved observations of UV-emitting stellar populations to assess the source of ionization for LI(N)ER-type gas.

## RESULTS

Using the WHAM (19, 20), we obtained high sensitivity ($I_{H_α} \sim 0.1\ R$), velocity-resolved ($R \sim 25{,}000$; $\Delta v \sim 12$ km s$^{-1}$) observations of several optical emission lines, e.g., Hα λ6563 Å, Hβ λ4861 Å, [NII] λ6584 Å, and [OIII] λ5007 Å, in the vicinity of Galactic Center. This dual-etalon Fabry-Perot spectrometer samples the region using a 1° diameter beam. These data are supplemented with HI 21-cm observations from the HI4PI survey (*22*), with a sensitivity of 43 mK and an angular resolution of 16.2 arcminutes.

[1]Department of Astronomy, University of Wisconsin-Madison, 475 N Charter St., Madison, WI 53706, USA. [2]Department of Physics, University of Wisconsin-Whitewater, 800 West Main Street, Whitewater, WI 53190, USA. [3]Department of Physical Sciences, Embry Riddle Aeronautical University, 1 Aerospace Blvd., Daytona Beach, FL 32114, USA. [4]Space Science Institute, 4750 Walnut St., Suite 205, Boulder, CO 80301, USA.
*Corresponding author. Email: krishnarao@astro.wisc.edu

In the low-extinction region around $(l,b) = (1°$ to $5°, -3°$ to $-6°)$, we detect Hα emission in the velocity range of $-110$ km s$^{-1} \leq v_{LSR} \leq -50$ km s$^{-1}$, the same velocity range in which 21-cm emission is seen. Figure 1 shows a map of HI 21-cm emission near Galactic Center for longitude $l > 0°$, with $-110$ km s$^{-1} \leq v_{LSR} \leq -50$ km s$^{-1}$ in blue and $l < 0°$ with $+50$ km s$^{-1} \leq v_{LSR} \leq +110$ km s$^{-1}$ in red. These "forbidden" velocities cannot arise from circular rotation in the inner Galaxy and have been interpreted as the expansion of a tilted circular annulus (*21*) or tilted elliptical trajectories of gas, possibly aligned with the bar (*23*). The contours on the HI 21-cm map show our detection of Hα emission in the same velocity windows. Two example spectra show kinematically distinct optical emission lines in the same velocity range as the neutral gas of the Tilted Disk. There is also Hα emission above the midplane of the Tilted Disk that we refer to as the "Upper Feature." Unlike the Tilted Disk emission, this emission is not kinematically distinct and is found in broad wings of local Hα emission. In some directions, we observe HI 21-cm emission at the same negative velocities as the Upper Feature, but this emission will not be considered further.

Figure 1 also shows that the [NII]/Hα emission line ratio associated with the Tilted Disk is substantially higher than the local emission traced by gas near $v_{LSR} = 0$ km s$^{-1}$. Our multiwavelength WHAM observations of [NII], [OIII], Hα, and Hβ allow for the inner regions of the Milky Way to be compared with other galaxies using the diagnostic Baldwin-Phillips-Terlevich (BPT) diagram (*6*). Figure 2 shows two optical line ratios in a diagram comparing Sloan Digital Sky Survey (SDSS) galaxies, WHAM observations of the Tilted Disk, and WHAM observations in the direction of the "Scutum Star Cloud", an extinction window at $l \sim 30°$, which traces gas in the Scutum spiral arm at $R_G$ = 4 to 6 kpc (24). Of the 10 Tilted Disk pointings shown on this plot (orange points and pink bars), 6 have only [NII]/Hα ratio measurements, 3 have upper limits for [OIII]/Hβ, and 1 pointing has both lines of the [OIII]/Hβ ratio detected at greater than 2σ significance. All but one of these pointings would be classified as composite or LI(N)ER emission. Moreover, the [NII]/Hα ratio is distinctly different from what is seen in HII regions, Galactic diffuse ionized gas in the disk, and the Scutum spiral arm but is similar to ratios seen in LI(N)ERs. Last, the face-on Hα surface brightness predicted by a geometric model of the ionized Tilted Disk (described in the next section) is $\sim 5 \times 10^{-17}$ ergs s$^{-1}$ cm$^{-2}$ arcsec$^{-2}$, comparable to the average surface brightness observed in M31 (25).

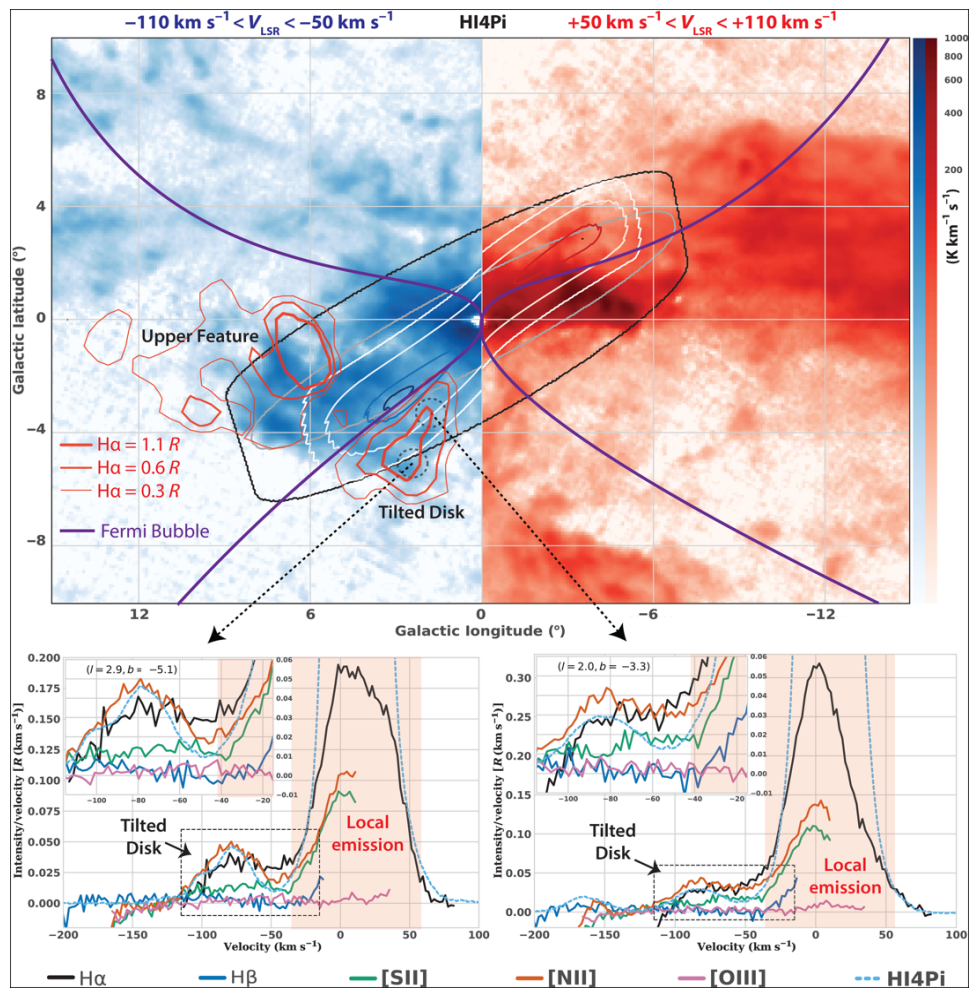

**Fig. 1. Observations of the Tilted Disk.** Integrated velocity channel map of HI 21-cm emission over inner Galaxy forbidden velocities for positive longitudes ($-110$ km s$^{-1} \leq v_{LSR} \leq -50$ km s$^{-1}$; bluescale) and negative longitudes ($+50$ km s$^{-1} \leq v_{LSR} \leq +110$ km s$^{-1}$; redscale), showing the tilted distribution of this gas. Hα contours integrated over the same velocity range show evidence for an ionized counterpart to the neutral gas. A projection of a tilted elliptical HI disk model (*23*) integrated over all velocities is shown with black and gray contours, while the blue-to-white and red-to-white contours show the emission predicted by the model in the same velocity range as the data with contour values of 0.1, 10, 75, and 100 K km$^{-1}$ s$^{-1}$. The purple line shows the projected outline of the Fermi Bubble (*38*) extrapolated into Galactic Center. The dotted circles and arrows show the location of two 1° WHAM beams and their corresponding optical emission line spectra for Hα, Hβ, [NII], [SII], and [OIII]. A rescaled (1/20) HI spectrum averaged over the WHAM beam is also shown to demonstrate the kinematic agreement between the neutral and ionized gas observations. The bright emission around $v_{LSR} = 0$ km s$^{-1}$ is from local emission in the solar neighborhood.

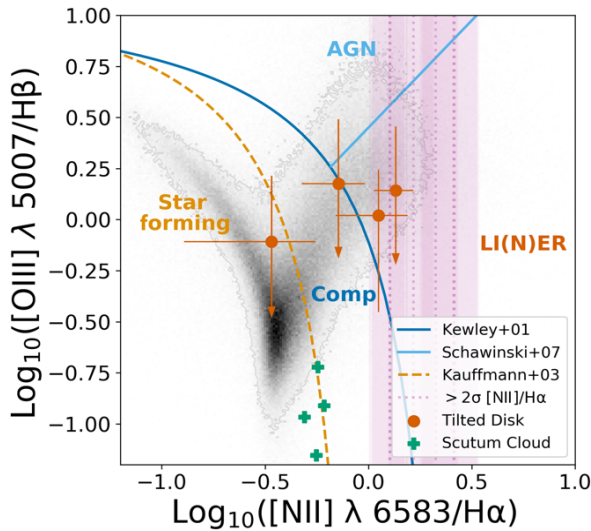

**Fig. 2. BPT diagram with the Milky Way.** BPT diagram of the [OIII]/Hβ versus [NII]/Hα line ratios for SDSS DR7 galaxies (grayscale) and the Milky Way (colored data), as observed with WHAM. Classification lines in blue, cyan, and orange separate regions by their modeled primary excitation mechanism (7–9). Orange-shaded filled circles are WHAM observations of the Tilted Disk structure we have modeled. Green plus symbols are extinction-corrected WHAM observations of the Scutum direction (24) and cover a range of 4 kpc < $R_G$ < 7 kpc. Pink lines and shaded regions show additional observations of [NII]/Hα of the Tilted Disk where either [OIII] or Hβ are not detected. Error bars are either 1σ errors or upper limits (arrows).

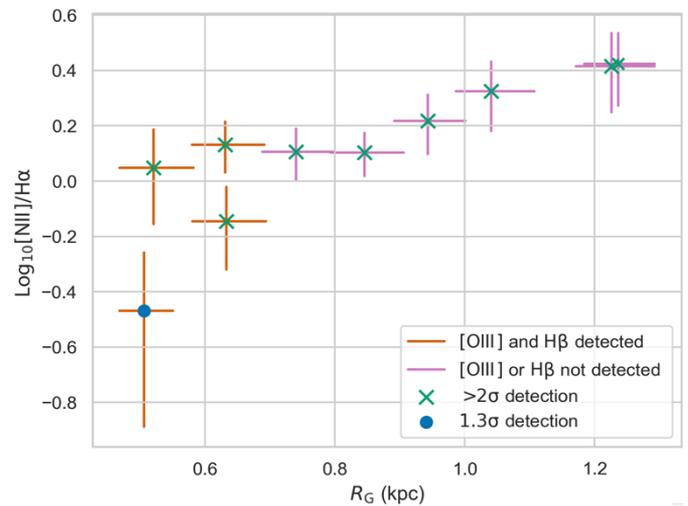

**Fig. 3. [NII]/Hα versus Galactocentric radius.** [NII]/Hα line ratio toward the Tilted Disk as a function of Galactocentric radius ($R_G$). Points with orange error bars correspond to the orange points in the BPT diagram of Fig. 2, while points with pink error bars correspond to the pink vertical bars in the BPT diagram. An additional offset along both axes of 0.01 is added to the point with the largest values on both axes to differentiate the two points with very similar values.

Using the same geometrical model to convert observed velocity to Galactocentric radius, Fig. 3 also shows evidence for a trend of increasing [NII]/Hα ratio with Galactocentric radius in the inner Galaxy. We also find that the spectra in the direction closest to Galactic Center are consistent with star formation–dominated spectra.

Interpreting these emission spectra further requires a three-dimensional (3D) kinematic model of gas to convert the observed gas velocity to the gas location within the Galaxy and to estimate the total ionized gas mass. We consider two different models: (i) a geometrical model of a tilted elliptical disk (23), modified to include an ionized gas component, and (ii) a hydrodynamical model of gas flow in the barred potential of the Milky Way (26) that includes heating, cooling, and a chemical network to track the ionization state and composition of the gas. Details for both models are given in the Supplementary Materials and are summarized here.

We first consider a previously constructed model of the neutral Tilted Disk by Liszt and Burton (23) and modify it to include an ionized gas component. Before the development of this model, anomalous velocity features were modeled as arising in explosive events at Galactic Center (27). The original Tilted Disk model (21) explained many of these observations as resulting from a tilted disk with radial expansion. Subsequent refinements of the model explained the anomalous velocity gas as elliptical motion requiring no expansion (23).

The neutral gas model has elliptical streamlines of gas, with a semimajor axis of $a_d'$ = 1.51 kpc and a semiminor axis of $b_d'$ = 0.49 kpc. The gas density varies only as a function of height; angular momentum is conserved along the elliptical orbits. The disk is tilted 13.5° out of the plane, 20° toward us, and has its major axis at an angle of 48.5° with respect to the Sun–Galactic Center direction (23). Our best-fitting ionized gas density model uses the same orientation and velocity field as the neutral gas and is characterized by a central midplane density $n_{e,0}$ = 0.39 (+0.06/−0.05) cm$^{-3}$ and central Gaussian scaleheight, $H_{z,0}$ = 0.26 ± 0.04 kpc. We also introduce a flaring factor, $F_z$ = 2.05 (+0.42/−0.31), such that the scaleheight increases linearly with radius as $H_z(r) = H_{z,0}[(1 - x) + F_z x]$ where $x = r/(a_d')$. Similar to the neutral gas model, we assume that surface density is independent of radius, so $n_e(r) = n_{e,0} H_{z,0}/H_z(r)$. As in previous work (28), we remove an inner ellipse with a semimajor/minor axes of $a_d'/2$ and $b_d'/2$, half the outer value.

This model predicts the observational trends of Hα seen with WHAM and confirms that directions where we do not detect emission are inaccessible due to extinction. The previous neutral gas model and our new ionized gas model allow us to compare the mass of these two components. We find a total neutral gas mass of $M_{H0}$ = $(3.1 \pm 0.3) \times 10^6$ solar masses, which agrees with previous estimates (28). The total ionized gas mass from our model is $M_{H+}$ = 12 (+4/−3) × $10^6$ solar masses. Since these values come from extrapolating our ionized gas model beyond the observational window where Hα is

**3 of 6**

detected, we also compare the total mass of neutral and ionized gas in our observing window. Over the region, $l = 0°$ to $6°$, $b = −7°$ to $−2°$ in the velocity interval, $v_{LSR} = −120$ to $−40$ km s$^{−1}$, and HI 21-cm observations combined with our geometric models yield a neutral and ionized gas mass of $(0.30 \pm 0.01) \times 10^6$ and $0.37$ $(+0.12/−0.09) \times 10^6$ solar masses, respectively, and an ionization fraction of 55% ($\pm 7\%$). Despite the uncertainties of how to extrapolate our results in the extinction window to the full structure, it is clear that a substantial fraction of the gas we observe is ionized. Details on the mass estimates and uncertainties shown here are available in the Supplementary Materials.

Although the geometric model has the advantage of simplicity, it does not capture the full complexity of the gas distribution expected in a barred galaxy like the Milky Way. In particular, hydrodynamical models show notable variations in density with both azimuth and radius interior to the bar radius [e.g., (*26*) and references therein]. From our vantage point at the Sun, looking in the direction of positive Galactic longitudes ($l > 0°$), these models predict that the leading—high positive $v_{LSR}$—side of the bar is characterized by dense gas and dust, while the trailing side of the bar—high negative $v_{LSR}$—would have much lower gas density. To examine the effects of this density asymmetry on our interpretation, the Supplementary Materials contains a comparison of our geometric model to a hydrodynamical simulation (*26*). Since this particular simulation also tracks the chemical state of the gas, it makes specific predictions for the density structure of ionized gas.

We find that this model fails to adequately account for the H$\alpha$ emission we see in three principal ways. First, it fails to predict the observed tilted distribution of gas, a shortcoming discussed in (*26*). Since the model gas layer lies in the plane, it lies behind large amounts of extinction, and we would expect to detect no H$\alpha$. Second, although the vertical thickness of the gas depends on position and is not easily characterizable with a single scaleheight, we find that at all positions, the model produces a gas layer much thinner than what is inferred from both the H$\alpha$ and HI observations. This issue was also noted by (*26*) with regard to the molecular and neutral atomic gas but is even more discrepant for the ionized gas component. Last, even when we introduced an arbitrary tilt in the gas distribution, we found that the gas on the trailing side of the bar, traced by negative-velocity gas, had a much lower density of ionized gas than needed to explain the observed H$\alpha$ emission. This last discrepancy is almost certainly due to the radiation field used in this simulation, which did not include hydrogen-ionizing photons. All of the hydrogen ionization in this model comes from cosmic rays or collisional ionization.

The discrepancies we identify are likely to be present in any of the currently available models of gas flow in a barred potential. To our knowledge, no model has successfully resulted in a tilted distribution. Moreover, in all models, the higher gravitational potential of the inner Galaxy can be expected to result in a thin gas layer in the absence of other sources of vertical support. While many models seem to produce a high positive-velocity dense gas structure on the leading side of the bar, analogous to the observed "Connecting Arm" seen in CO observations, there has been much less attention paid to matching the column densities of gas in the negative-velocity trailing side of the bar. Regardless, both our geometrical model and the hydrodynamical models agree on the general location within the Galaxy where the kinematic signatures we observe must originate—the trailing side of the bar.

## DISCUSSION

Given that the presence of LI(N)ER-type gas has been shown to be well-correlated with the occurrence of Galactic bars (*29*), one might have reasonably suspected the ionized gas in the inner Milky Way to have LI(N)ER-like line ratios. Our observations not only confirm this expectation but allow for new constraints on the nature of the ionization mechanism and the power requirements. Particularly intriguing is our finding that the [NII]/H$\alpha$ ratio drops as the Galactocentric radius decreases, with our innermost observations consistent with a star formation–dominated ionizing radiation field. Analysis of far-infrared line ratios in the CMZ, which are not accessible optically, has been interpreted as a star formation–dominated radiation field (*30*), although the situation is not entirely clear (*31*). With the caveat that the true density structure of ionized gas is likely to be more complicated than assumed in our geometric model, we can use our H$\alpha$ observations to infer the level of ionizing flux in the inner Galaxy. If one assumes that the disk is ionized from outside, then a balance between the number of ionizations and recombinations in our model exponential ionized disk implies a (plane-parallel) hydrogen-ionizing flux of $\phi = (2.7 \times 10^7$ photons s$^{−1}$ cm$^{−2}$) $(1 - 0.5 x)$ to each side of the disk, where $x = r/a_d'$. This is 10 times the flux needed to maintain the warm ionized layer in the vicinity of the Sun (*32*). The corresponding luminosity of hydrogen-ionizing photons is $Q(H^0) = 6.3 \times 10^{50}$ photons s$^{−1}$. This can be compared to the number of Lyman continuum photons coming from star formation in the CMZ, which is $Q_{CMZ} = 1.2$ to $3.5 \times 10^{52}$ photons s$^{−1}$, assuming a CMZ star formation rate of 0.05 to 0.15 solar mass year$^{−1}$ (*33*, *34*) and that a star formation rate of 1 solar mass year$^{−1}$ produces $2.4 \times 10^{53}$ photons s$^{−1}$ (*35*).

Although it would only take approximately 5 to 10% of these ionizing photons to maintain the ionization in the Tilted Disk, models of radiative transfer will be needed to establish whether this radiation source alone would suffice to explain the changing level of ionization seen. Evolved stellar populations, e.g., subdwarf OB stars, have been posited as a source of ionization in LI(N)ER systems; the contribution of these sources could be observationally constrained in the Galaxy. A Galactic bulge concentration of these stars would occur at magnitude $m_V \sim 19$ for a subdwarf OB absolute magnitude of $M_V = 4.6$ (*36*). Since the Galactic bulge is vertically thicker than much of the extinction in the inner Galaxy, such a population could be detected.

Absorption line studies will also be valuable for studying the thermal pressure, ionization, metallicity, and dust depletion in the neutral and ionized components of this LI(N)ER-like gas. One such study toward the distant B1 Ib-II star LS 4825 ($l = 1.67°$ and $b = −6.63°$) shows absorption at the same velocity as the Tilted Disk. Analysis of this component yields a solar abundance of sulfur, ~1-dex depletion of iron and aluminum, a thermal pressure three times higher than in the solar neighborhood, and evidence for more highly ionized gas traced by C IV and Si IV, indicating a possible interface with hotter gas (*37*).

Last, hot gas associated with a large-scale nuclear outflow indicated by the Fermi Bubbles may provide another source of ionizing radiation (*38*). In Fig. 1, we show how the estimated boundary of the Fermi Bubble—as seen in gamma-ray and x-ray emission—compares to the structure of the Tilted Disk. If hot gas is intermixed with the Tilted Disk structure, then our gas mass estimates based on H$\alpha$ emission will be lower limits (*39*) to the total amount of ionized gas in the inner Galaxy. Our observations show that the nearest

LI(N)ER-like gas to us in the universe is now the inner Milky Way and no longer the inner parts of M31. This opens new avenues to better constrain the nature and ionization sources of this elusive class of gas with an unprecedented level of detail across all wavelengths.

## MATERIALS AND METHODS

All data used in this work can be accessed publicly. The WHAM Sky Survey has a public release of H$\alpha$ observations available online at http://www.astro.wisc.edu/wham-site/. The HI4PI observations of 21-cm neutral hydrogen (*22*) and the 3D dust models (*40*) used in this work are also available publicly. The Tilted Disk geometric model can be computed using the open-source Python package, modspectra (https://github.com/Deech08/modspectra). Multiwavelength WHAM observations have not been publicly released, but those specific to this work can be accessed at the following GitHub repository, which also includes Python notebooks to replicate plots shown in this work (https://github.com/Deech08/mw_bpt).

### Statistical analysis

For each WHAM pointing, the H$\alpha$ emission intensity and line profiles are calculated for our model assuming Case B recombination and using maps of the 3D dust distribution to apply extinction corrections (*40*). A Bayesian Markov Chain Monte Carlo approach is used to optimize the model in comparison with WHAM observations of H$\alpha$ in the extinction windows shown in Fig. 1. Our model predicts the observational trends of H$\alpha$ seen with WHAM and confirms that directions where we do not detect emission are inaccessible due to extinction. Full details on this procedure are available in the Supplementary Materials.

## SUPPLEMENTARY MATERIALS

Supplementary material for this article is available at http://advances.sciencemag.org/cgi/content/full/6/27/eaay9711/DC1 and appended after this article.

**Acknowledgments:** This work benefited from discussion with F. J. Lockman, N. McClure-Griffiths, B. P. Wakker, A. J. Fox, and L. D. Anderson. We thank M. Sormani for providing access to and insight on the hydrodynamical simulations used as a reference model. **Funding:** We acknowledge the support of the NSF for WHAM development, operations, and science activities. The survey observations and work presented here were funded by NSF awards AST-0607512, AST-1108911, and AST-1714472/1715623. R.A.B. would like to acknowledge support from NASA grant NNX17AJ27G. D.K. would like to acknowledge support from the J.D. Fluno Family Distinguished Graduate Fellowship. Some of this work took part under the program Milky Way Gaia of the PSI2 project funded by the IDEX Paris-Saclay, ANR-11-IDEX-0003-02. **Author contributions:** D.K. led the investigation, formal analysis, methodology, software, and visualization. R.A.B. led the conceptualization and project administration, and contributed to funding acquisition and methodology. L.M.H. led funding acquisition and resources, and contributed to project management and software. D.K. and L.M.H contributed equally to data curation; supervision was equally conducted by R.A.B. and L.M.H.; D.K., R.A.B., and L.M.H. equally contributed to validation, writing of the first draft, reviewing and editing. **Competing interests:** The authors declare that they have no competing interests. **Data and materials availability:** All data needed to evaluate the conclusions in the paper are present in the paper, the Supplementary Materials, and/or the associated GitHub repository. Additional data related to this paper may be requested from the authors.

Submitted 1 August 2019
Accepted 7 April 2020
Published 3 July 2020
10.1126/sciadv.aay9711

**Citation:** D. Krishnarao, R. A. Benjamin, L. M. Haffner, Discovery of diffuse optical emission lines from the inner Galaxy: Evidence for LI(N)ER-like gas. *Sci. Adv.* **6**, eaay9711 (2020).

**Supplementary Materials**

Fig. S1. Tilted Disk Model Schematic.
Fig. S2. Modeled and Observed Hα Spectra.
Fig. S3. Ionized Tilted Disk Posterior Distributions.
Fig. S4. Predicted Tilted Disk Hα Map.
Fig. S5. Predicted Hydrodynamic Face-On Hα Map.
Fig. S6. Edge-on Hα Comparisons.
Fig. S7. Edge-on HI Comparisons.
Table S1. Tilted Disk Model Parameters.
Table S2. Comparison of Masses and Vertical Extent of Models.

**Optical Emission Lines**

The Hα data were taken as a part of the WHAM Sky Survey (WHAM-SS) *(19, 20)*. Each 30 second observation obtains a 200 km s$^{-1}$ velocity-range spectrum around Hα integrated over a 1° beam. The dataset presented here toward the Tilted Disk is derived primarily from the southern portion of the survey with WHAM sited at Cerro Tololo. While similar to those released in WHAM-SS DR1, these spectra provide a more calibrated view around the nuclear region. They will be fully integrated into the DR2 release and the survey can be accessed with the open-source python package, *whampy (41)* (see the WHAM-SS release documentation for details: http://www.astro.wisc.edu/wham/). Other emission-line data have been obtained as part of ongoing multi-wavelength WHAM surveys using longer (60s) observations. These spectra are processed in the same way as Hα by applying a flat-field, subtracting an atmospheric template, and subtracting a constant baseline to reach a 3σ sensitivity of 0.1 R.

**Geometric Model: Neutral Gas**

The original HI model was designed to describe the velocity field of neutral gas in the inner Galaxy, starting with a circular model *(21, 42)* and moving on the elliptical model we consider *(23)*. Future iterations added additional gas tracers, and included the Central Molecular Zone *(43, 44)*. We verified that this model still provides an adequate fit to the modern HI4PI survey data *(22)* and take this model "as is" but update the distance to Galactic Center from 10 kpc to 8.127 kpc *(45)*. HI observations towards the inner Galaxy are most sensitive to the velocity field of the gas, as opposed to density structures, because of the large velocity gradients in this environment. Because of this, mass estimates made using the HI model may have large errors.

The iso-density contours of the neutral gas model *(23)* follows ellipses of the form

$$\frac{x_d^2}{a_d^2} + \frac{y_d^2}{b_d^2} = 1$$

where $a_d$ and $b_d$ are the semi-major and semi-minor axes, respectively, and

$$a_d = b_d \left( 1.6 + 1.5 \frac{b_d}{b_d'} \right)$$

with $b_d' = 0.488$ kpc. The tangential velocity along the minor axis depends only on $b_d$, with the form

$$v_t(b_d) = 360 \text{ km s}^{-1} \left[ 1 - \exp\left( \frac{-b_d}{0.1 \text{ kpc}} \right) \right]$$

and assumes angular momentum conservation along the elliptical path. Coordinates from the elliptical disk frame ($x_d$, $y_d$, $z_d$) are transformed to the Galactocentric frame ($x$, $y$, $z$) using

$$\begin{bmatrix} x \\ y \\ z \end{bmatrix} = \begin{bmatrix} \cos\beta \cos\theta & \cos\beta \sin\theta & -\sin\beta \\ \cos\theta \sin\alpha \sin\beta - \cos\alpha \sin\theta & \cos\alpha \cos\theta + \sin\alpha \sin\beta \sin\theta & \cos\beta \sin\alpha \\ \cos\alpha \cos\theta \sin\beta + \sin\alpha \sin\theta & \cos\alpha \sin\beta \sin\theta - \cos\theta \sin\alpha & \cos\alpha \cos\beta \end{bmatrix} \begin{bmatrix} x_d \\ y_d \\ z_d \end{bmatrix}$$

where $\alpha = 13.5°$ describes the tilt of the disk about the x-axis, $\beta = 20° = 90° - i$ describes the inclination, $i$, of the disk, and $\theta = 48.5°$ describes the angle between the major axis of the elliptical disk and the x-axis. Although $\theta$ does not line up with current estimates of the bar/bulge angle, for the purposes of estimating the relative amounts of neutral and ionized gas, we chose not to modify this parameter.

In this picture, the origin is at Galactic Center, the positive x-axis points parallel to the sun-Galactic Center direction, and the positive y-axis points parallel to the $l = 90°$ direction. $\beta = 0°$ corresponds to an edge-on disk. $\alpha$, $\beta$, and $\theta$ describe rotations along the x, y, and z axes, respectively. Positive $\alpha$ indicates that positive longitudes are rotated below the Galactic plane; positive $\beta$ indicates that the near side of the disk is rotated below the Galactic plane; positive $\theta$ indicates clockwise rotation as viewed from above.

Radial velocities are converted to local standard of rest (LSR) velocities when computing the model. HI emission within each cell is calculated from the gas density as follows. The emission along a single line of sight at some LSR velocity, $v$, is given by

$$T_b(v) = T_{gas}\left( 1 - e^{-\tau(v)} \right)$$

where $T_b(v)$ is the brightness temperature observed at some velocity channel, $T_{gas}$ is the temperature of the neutral hydrogen gas, and $\tau(v)$ is the optical depth at some velocity channel. The optical depth is computed using

$$\tau(v) = \sum_i \Delta\tau_i(v)$$

$$\Delta\tau_i(v) = n_{H_i} \frac{33.52}{T_{gas}\, \sigma_i} \exp\left(-\frac{1}{2}\left[\frac{v-v_i}{\sigma_i}\right]^2\right)\left(\frac{\Delta d_i}{50\ \text{pc}}\right)$$

where $\Delta\tau_i(v)$ is the optical depth of neutral hydrogen gas at some velocity, $v$, within cell $i$, $n_{H,\,i}$ is the neutral hydrogen gas density within the cell, and $\sigma_i$ is the gas velocity dispersion within the cell, $v_i$ is the radial LSR velocity of the neutral hydrogen gas within the cell, and $\Delta d_i$ is the width of the cell in parsecs *(46)*. For this model, $T_{gas} = 120$ K and $\sigma_i = 9$ km s$^{-1}$. We have compared the results of our model to longitude-velocity slices at fixed latitude and found good agreement. A synthetic HI datacube can be computed using *modspectra* via *cube.EmissionCube.create_LB80()*.

We also provide estimates of both the total neutral mass and the mass of neutral gas in the same direction and velocity range as where we detect ionized gas. The original neutral Tilted Disk model did not provide uncertainty estimates, so we adopt an uncertainty of 30%. Near Baade's window, we estimate the neutral gas mass using the HI 21-cm observations integrated over the directions $l = 0°$ to $6°$, $b = -7°$ to $-2°$ within the velocity interval, $v_{LSR} = -120$ to $-40$ km s$^{-1}$, and using distance estimates from the Tilted Disk model. Assuming optically thin media *(47)*, we estimate the neutral gas mass as

$$M_{H^0} = \sum_i N_{HI,i}\left(D_i\, \Delta\theta\right)^2$$

$$N_{HI,i} = \int_{-120\ \text{km s}^{-1}}^{-40\ \text{km s}^{-1}} 1.82\times 10^{18}\ \text{cm}^{-2}\text{K}^{-1}\text{km}^{-1}\text{s}\ T_i(v)\, dv$$

where $i$ refers to each observed pixel in the HI4PI data, $\Delta\theta$ is the angular size of the pixel in radians, $T_i$ is the observed brightness temperature in K, and $D_i$ is the distance to emitting gas taken from the Titled Disk model. Our uncertainties in the observed neutral gas mass in the direction of Baade's window are primarily propagated from errors in the model distance estimates.

### Geometric Model: Ionized Gas

We adopt the orientation and kinematics of the neutral gas model described above, but vary the density structure of ionized gas and include the extinction along the line of sight in order to predict the Hα profile. The ionized gas density is modeled as a function of

$$n_e(r, z_d) = n_{e,r}(r)\exp\left[-\frac{1}{2}\left(\frac{z_d}{H_{z,r}(r)}\right)^2\right]$$

$$x = r/r_{max}$$

radius and height as

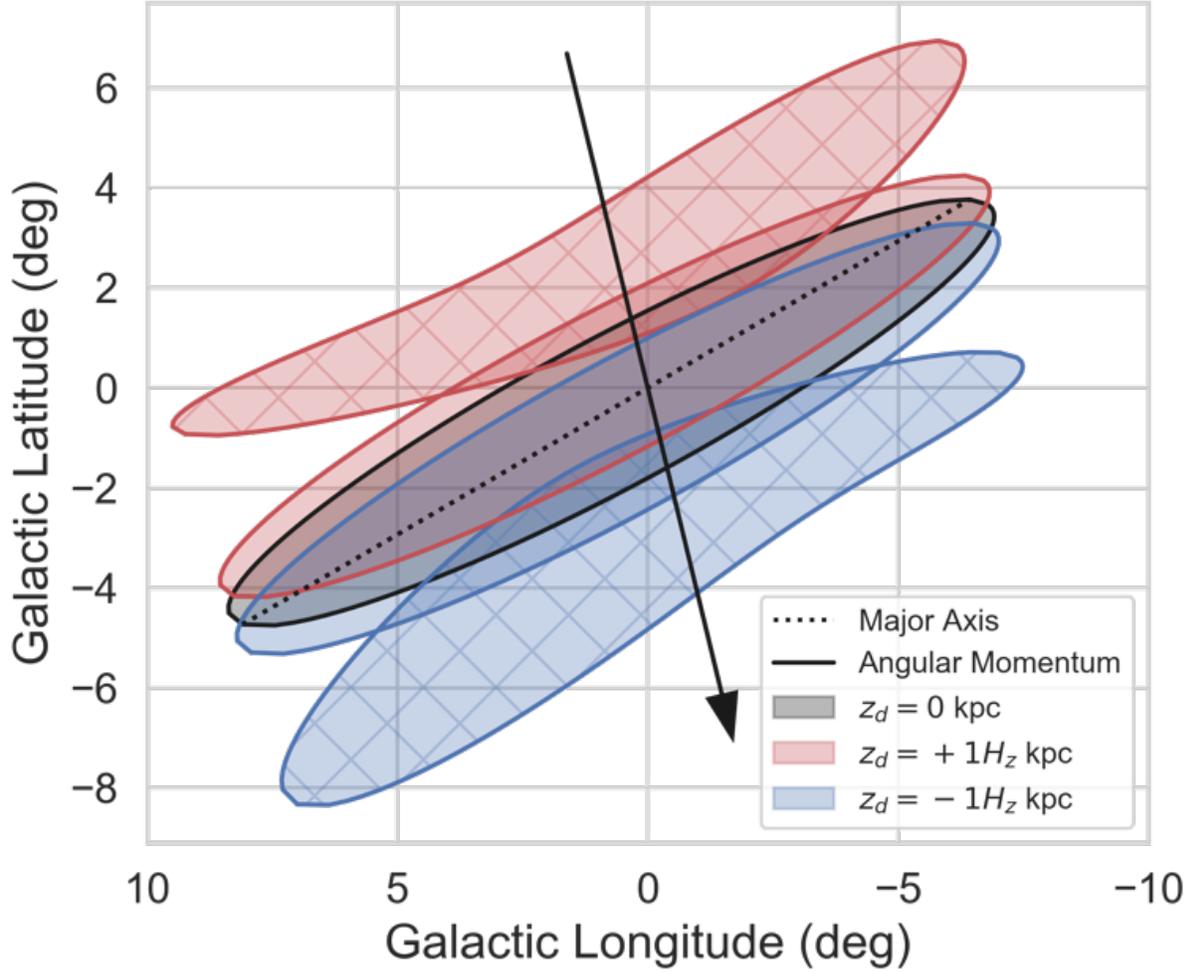

**Fig. S1. Tilted Disk Model Schematic.** A schematic of the updated tilted elliptical model of neutral (shaded) and ionized gas (shaded and hatched), showing the disk midplane (grey), and disk at $z_d = \pm H_z$ (red/blue) projected onto the sky in Galactic Coordinates. An angular momentum vector is shown going through the center of the disk and the major axis of the ellipse along the midplane is shown with a dotted line.

where $r$ and $z_d$ are in cylindrical coordinates for the tilted disk, $n_{e,r}(r)$ is the midplane ionized gas density as a function of $r$, and $H_{z,r}(r)$ is the vertical scale height of the ionized gas as a function of $r$. Both $n_{e,r}(r)$ and $H_{z,r}(r)$ depend on the amount of flaring as parameterized by a flaring factor $F_z$ such that

$$n_e(r) = n_{e,0} H_{z,0} / H_z(r)$$

where $r_{max}$ is the max radial coordinate of the disk, corresponding to the value of the largest semi-major axis ($a_d' = 1.5128$ kpc), and $n_{e,0}$ and $H_{z,0}$ are the midplane ionized gas density and ionized gas scale height at $r = 0$, respectively. **Figure S1** shows a schematic

$$H_z(r) = H_{z,0}[(1-x) + F_z x]$$

of the model Tilted Disk projected on the sky, showing the disk midplane (grey) and ±1 scale height (red/blue) along with an angular momentum vector.

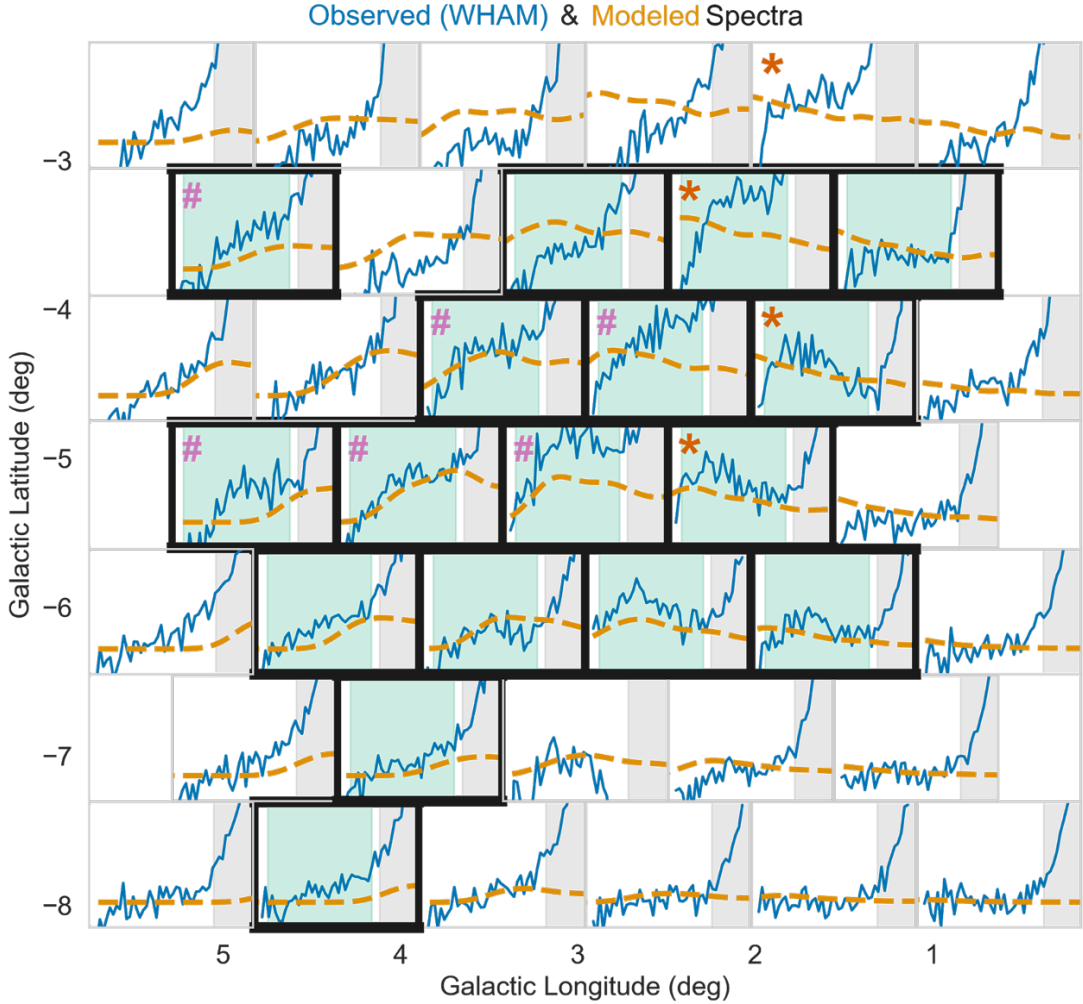

**Fig. S2. Modeled and Observed Hα Spectra.** Map of Hα Emission from WHAM (blue solid line) and synthetic observations (orange dashed line) of the ionized tilted disk towards Baade's Window; spectra are shown between -150 km s$^{-1}$ < $v_{LSR}$ < -20 km s$^{-1}$. Spectra outlined in black are used in the ionized gas model fitting process and are selected based on having a mean velocity in the "forbidden" range associated with the tilted disk (as opposed to broad wings from local emission) and sufficient Hα emission. The green shaded regions show the velocity range considered when optimizing the model and the grey shaded regions show local emission not considered in this model. Orange asterisks mark the directions where we are able to place points on the BPT Diagram in **Fig. 2.** Pink pound symbols mark the six directions where only [NII]/Hα is detected; on the BPT diagram of **Fig. 2**, these are indicated with vertical lines.

The Hα emission in photon units of Rayleighs (R) is calculated for our model with Case B recombination using

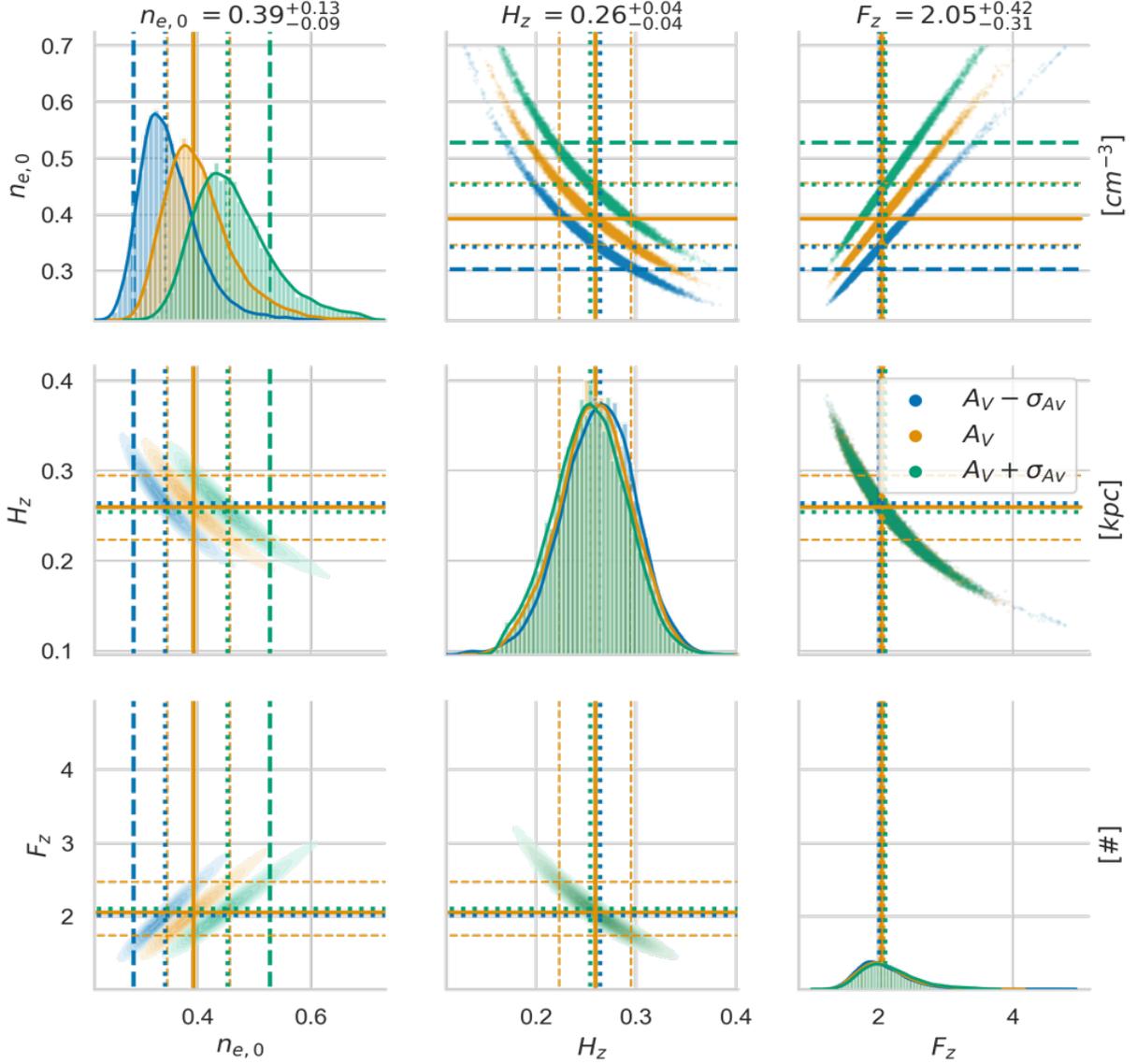

**Fig. S3. Ionized Tilted Disk Posterior Distributions.** Posterior distributions of the ionized gas model parameters, $n_{e,0}$, $H_{n,0}$, and $F_z$ for three different model fitting runs for the mean estimated $A_V$ (orange) and upper (blue) and lower (green) $A_V$ estimates. The solid line shows the 50th percentile of the mean $A_V$ posterior distribution and the 16th and 84th percentiles are shown with dashed lines as estimates for the uncertainties in these parameters. Extinction based uncertainties only strongly affect the $n_{e,0}$ parameter and our quoted uncertainty for this parameter is the 16th and 84th percentile from the low and high extinction runs, respectively. Dotted lines show the 50th percentile of the posterior distribution for the upper and lower $A_V$ estimates. Our adopted uncertainties correspond to the dashed orange lines for $H_{n,0}$, and $F_z$, and the dashed green and blue lines for $n_{e,0}$. The density parameter is the only one strongly affected by different extinction estimates.

| Neutral Gas Model | | | |
|---|---|---|---|
| Parameter | Symbol | Value | Units |
| Semi-Minor Axis | $b_d'$ | 0.488 | kpc |
| Semi-Major Axis | $a_d'$ | 1.5128 | kpc |
| Max Tangential Velocity | $v_{T,max}$ | 360 | km s$^{-1}$ |
| Tilt Angle | $\alpha$ | 13.5 | degrees |
| Inclination | $i$ | 70 | degrees |
| 90° - Inclination | $\beta$ | 20 | degrees |
| Major Axis Angle | $\theta$ | 48.5 | degrees |
| Vertical Scaleheight | $h_d$ | 81.27 | pc |
| Midplane Gas Density | $n_0$ | 0.33 | cm$^{-3}$ |
| Gas Temperature | $T_{gas}$ | 120 | K |
| Gas Velocity Dispersion | $\sigma_{gas}$ | 9 | km s$^{-1}$ |
| **Ionized Gas Model** | | | |
| Parameter | Symbol | Value | Units |
| Vertical Scaleheight | $H_z$ | $0.26 \pm 0.04$ | kpc |
| Vertical Flaring Factor | $F_z$ | $2.05^{+0.42}_{-0.31}$ | None |
| Midplane Gas Density | $n_{e,0}$ | $0.39^{+0.06}_{-0.05}$ | cm$^{-3}$ |
| Gas Temperature | $T_e$ | 8000 | K |
| Gas Velocity Dispersion | $\sigma_e$ | 12 | km s$^{-1}$ |

**Table S1. Tilted Disk Model Parameters.** Summary of neutral gas model parameters and results adopted from *(23)* and the ionized gas model from this work.

$$I_{H\alpha}(v) = \left(0.1442 \frac{R}{km\ s^{-1}}\right) \sum_i \left(\frac{n_{e,i}}{cm^{-3}}\right)^2 \left(\frac{\Delta d_i}{pc}\right)$$

$$\times \left(\frac{\sigma_i}{km\ s^{-1}}\right)^{-1} (T_{e,i}/10^4\ K)^{b_\lambda}$$

$$\times \exp\left(-\frac{1}{2}\left[\frac{v-v_i}{\sigma_i}\right]^2\right) e^{-\tau_\lambda}$$

where $n_{e,i}$ is the electron density within the cell, $T_{e,i}$ is the electron gas temperature, $b_\lambda = -0.942 - 0.031 \ln(T_{e,i}/10^4\ K)$ is from the case B recombination rate for H$\alpha$ *(48)*, and $e^{-\tau_\lambda}$ is the dust attenuation factor with $e^{-\tau_\lambda} = 10^{-1/2.5\ A\lambda}$. We use $A_K$ from the 2MASS-based three-dimensional extinction model of Marshall et al. *(40)*, converting this into $A_{H??}$ using $R_V = 3.1$ and the extinction curve of Fitzpatrick & Massa *(49)*. The extinction values are queried using the open-source python package, *dustmaps (50)*. We use these infrared-based maps since they allow us to obtain the extinction out to the distance of Galactic Center. The combination of WHAM H$\alpha$ and H$\beta$ emission offers an alternate way to estimate the extinction, but have the drawback that features at different distances can blend at the same velocity. Our model cell size is 0.1° x 0.1° in solid angle, whereas the three-dimensional dust model provides extinction estimates in 0.25° x 0.25° square cells. These are convolved with the 1° beam size of WHAM observations.

For comparison with the neutral gas, we calculate the total mass of ionized gas and the ionized gas mass in the extinction window where H$\alpha$ is detected. Errors in the ionized gas mass are estimated from the posterior distribution of parameters.

For our Bayesian parameter estimation implemented via the open-source python package, *emcee (51)*, we use Gaussian priors of the form

$$\ln p(\theta) \sim -\frac{1}{2}\left[\left(\frac{n_{e,0} - 0.3\ cm^{-3}}{0.25\ cm^{-3}}\right)^2 + \left(\frac{H_{z,0} - 1\ kpc}{0.5\ kpc}\right)^2 + \left(\frac{F_z - 2}{3}\right)^2\right].$$

The priors also provide a constraint forcing $n_{e,o} > 0$, $H_{z,0} > 0$, and $F_z \geq 1$. This biases our posterior distribution but prevents non-physical results or a scale height that decreases as a function of $r$. Our priors are chosen based on the behavior of more local and global Milky Way ionized gas *(52, 53)* and comparisons with previous HI model modifications that include some flaring *(43, 44)*. The likelihood function uses the Gaussian errors estimated for each WHAM observation as a function of velocity $\sigma_{H\alpha,\ WHAM}$ and has the form

$$\ln p(model|\theta) \sim -\frac{1}{2} \sum_v \frac{I_{H\alpha,model}(v,\theta) - I_{H\alpha,WHAM}(v)}{\sigma^2_{H\alpha,WHAM}(v)}$$

where $I_{H\alpha,\ WHAM}(v)$ is the WHAM H$\alpha$ observation with the local emission subtracted out as a single Gaussian (usually near $v_{LSR} \sim 0$ km s$^{-1}$) and $I_{H\alpha,\ model}(v, \theta)$ is the H$\alpha$ emission predicted by our ionized gas model with parameters $\theta$ using cells from a 200×200×200 grid that lie within a 1° WHAM beam. Additionally, $I_{H??,\ WHAM}(v)$ is limited to $v_{LSR} < -35$ km s$^{-1}$ to focus on the "forbidden" velocity gas.

**Figure S2** shows observed and synthetic Hα observations towards Baade's Window. A total of 17 WHAM beams surrounding the region of Baade's window are used for the model fitting. These points are selected based on their Hα emission signal strength and intensity-weighted peak velocity with the criteria of $I_{H??} > 0.1$ R and -90 km s$^{-1}$ < $v_{H??}$ < -55 km s$^{-1}$. These 17 pointings are marked in **Fig. S2** with black outlines. The results of the Bayesian parameter estimation are shown in **Fig. S3**. **Table S1** contains the resulting parameters for both the neutral and ionized gas models.

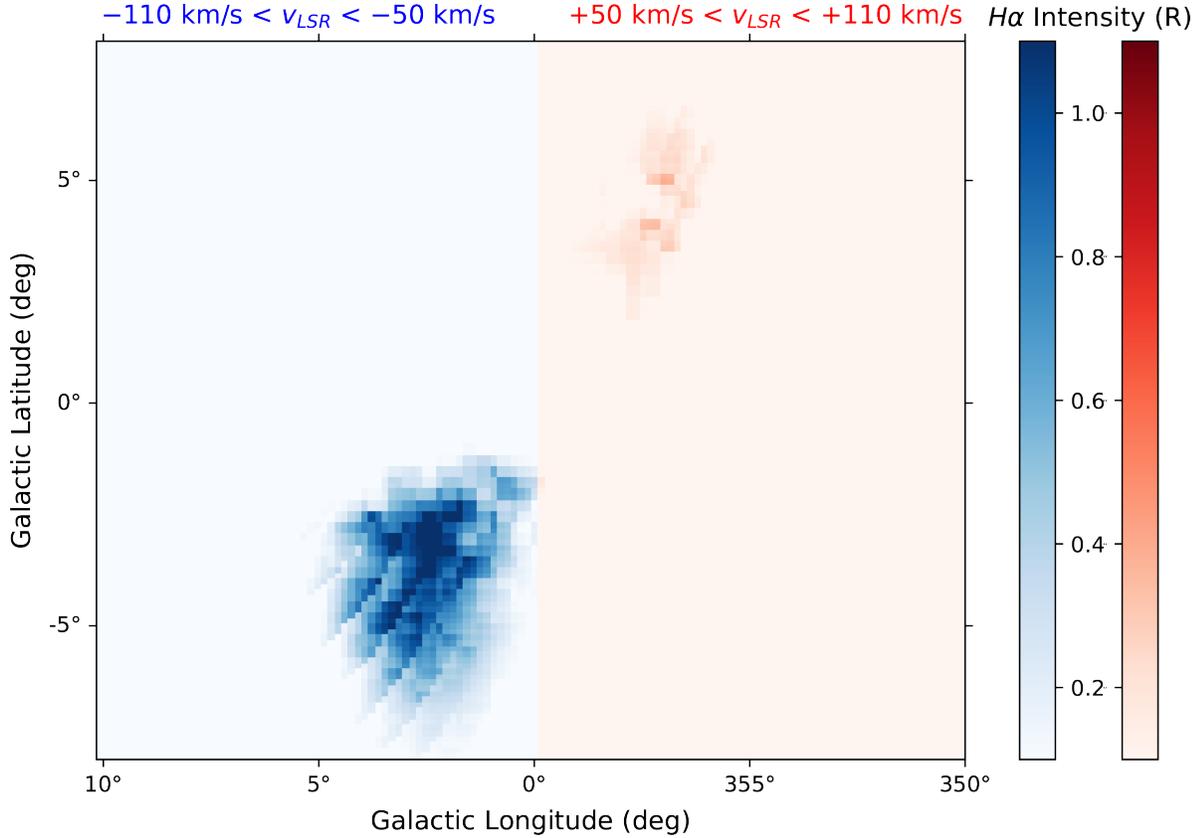

**Fig. S4. Predicted Tilted Disk Hα Map.** An integrated Hα map at the same velocities as in **Fig. 1** from the Tilted Disk model showing that we only expect to see detectable emission through the area surrounding Baade's low extinction window. The red emission at positive velocities above the plane is very faint and not currently detected in WHAM observations.

We run our MCMC parameter estimation three times, using a mean $A_V$ and ±1 $\sigma$ errors as provided in the 3D dust map *(40)*. Changing our $A_V$ estimate only significantly changes the posterior distribution of the $n_{e,0}$ parameter. Our quoted parameters are the 50th percentile with uncertainties estimated using the 16th and 84th percentiles from the posterior distribution of the mean $A_V$ MCMC run. Our uncertainties for the density parameter $n_{e,0}$ are computed using the 16th percentile of the $A_V$ - $\sigma$ and 84th percentile of the $A_V$ + $\sigma$ MCMC run. A synthetic Hα data cube can be computed using *modspectra* via *cube.EmissionCube.create_DK20()*.

In our final adopted model of ionized gas, we find that extinction is high enough over most of the structure that along a majority of sightlines the Hα should be undetectable. Our model correctly predicts, as shown in **Fig. S4**, that we should only detect ionized gas in the windows where we do, in fact, see it.

**Comparison with Hydrodynamical Models**

Our reference HI model (*23*) is chosen because of its simplicity and its ability to account for the tilted distribution of gas in the inner Galaxy. However, this model does not include any density variations with azimuth or radius as is seen in modern simulations and observations of other galaxies. A complete understanding of the gas in this region requires a more hydrodynamical approach to explain asymmetries in the vicinity of the CMZ, as demonstrated in (*26*). These simulations, designed to explain gas flowing in a Milky Way-like bar potential, include a photo-dissociating interstellar radiation field and a chemical network with adiabatic cooling dependent on the chemical composition. A snapshot of this simulation at 181 Myr provides a qualitative match to the observed CO longitude-velocity features observed in the inner Galaxy. These simulations provide predictions for the distribution and physical conditions of the multiphase ISM and are used here to compare with HI and Hα observations and the neutral and ionized Tilted Disk model.

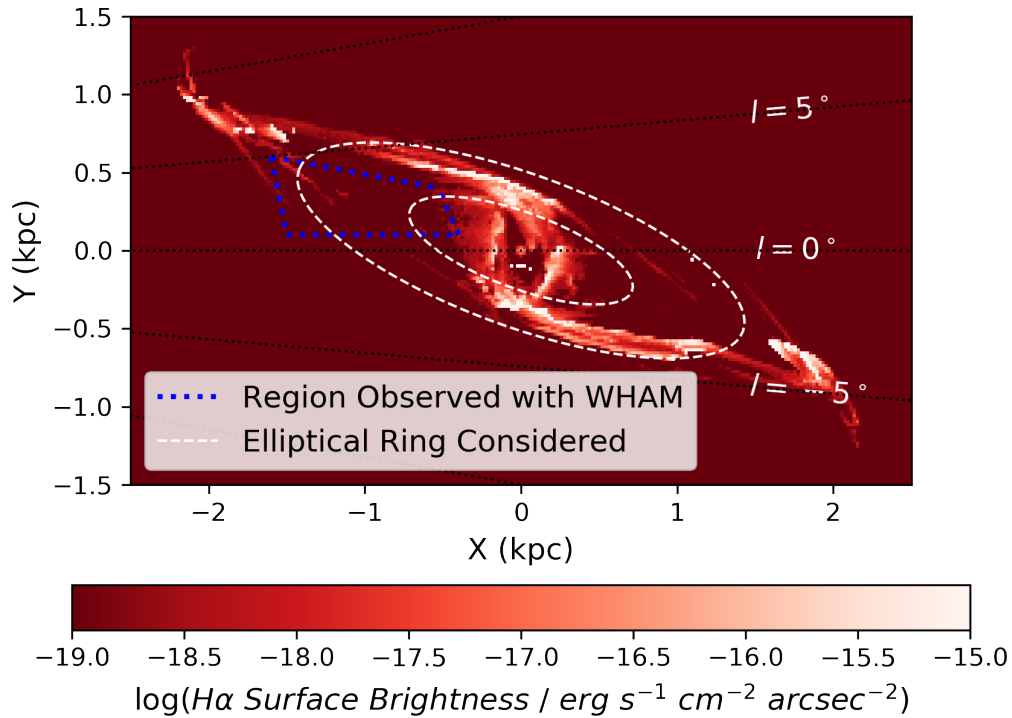

**Fig. S5. Predicted Hydrodynamic Face-On Hα Map.** Face-on map of Hα surface brightness from hydrodynamical simulations of (*26*) for t=181 Myr. Blue dotted lines enclose the region of negative-velocity gas that would be probed by WHAM observations and white dashed lines enclose an ellipse with the same physical scale as the Tilted Disk model rotated in angle to align with the simulations. The midplane density of ionized gas in the region probed by WHAM observations is approximately $n_{H+}=10^{-4}$ cm$^{-3}$.

**Fig. S5** shows a predicted face-on distribution of Hα surface brightness in the inner region of a t=181 Myr snapshot from our comparison hydrodynamical model (*26*). The negative velocity emission observed with WHAM would originate from within the dashed blue outlines on this figure. However, these simulations predict a very low density of both neutral and ionized gas in this region; most of the mass for the l>0° direction is concentrated in the positive-velocity dense "dust lane" features. In **Fig. S6**, we show the Hα emission prediction for both the hydrodynamical model (left column) and the modified Tilted Disk model (right column) viewed edge-on from the position of the Sun, shown both with (bottom row) and without (top row) extinction. Emission from the hydrodynamical model subtends a thin layer, which would be unobservable given the extinction in the midplane. In the central column, we artificially tilted the hydrodynamical simulations in order to align the disk midplane with our extinction window. A comparison with our modified Tilted Disk model (right column) shows that the hydrodynamical models produce a much thinner and fainter ionized gas structure.

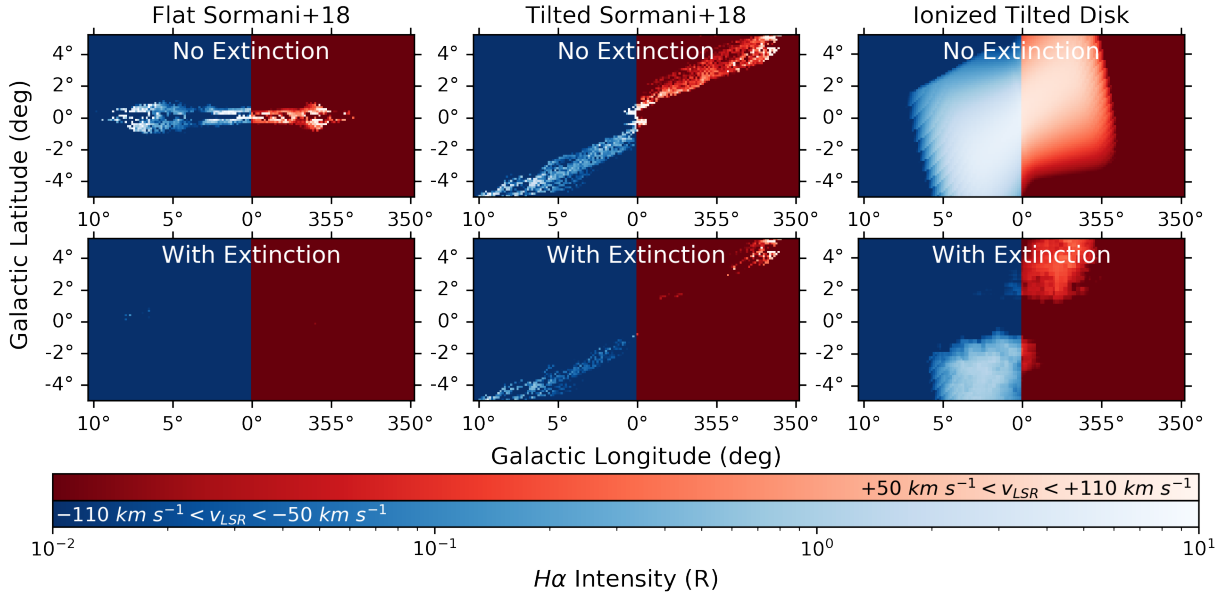

**Fig. S6. Edge-on Hα Comparisons.** Edge-on maps of Hα intensity from hydrodynamical simulations of (*26*) and the Tilted Disk from this work. The top row shows projections with no extinction, while the bottom row shows the same projections, but with extinction from the 3D dustmaps of (*40*). Maps are restricted to the "forbidden" velocities as shown within the colorbars. The left column shows the hydrodynamical simulations as originally orientated in the Galactic plane (*26*); the middle column shows the same simulations, but tilted to the same position angle as the Tilted Disk; the right column shows the ionized Tilted Disk from this work. The hydrodynamical models, even when artificially tilted, produce a layer that is much thinner and fainter than the best fitting Tilted Disk model.

Both the thinness of the molecular and atomic gas layer in these simulations and the lack of a tilt were discussed by the developers of this model (*26; see their section 5.5*). The thinness is attributed to the lack of stellar feedback and star formation in the simulations. Including these effects could provide additional turbulent pressure support. **Fig. S7** provides comparison of the predicted HI column density for the hydrodynamic model— both flat and tilted—with the geometrical Tilted Disk model and the observations. This demonstrates that the Tilted Disk model provides a better representation of the negative velocity gas that is the focus of this paper. For this reason, we adopt this model in order to infer the physical parameters of this inner Galaxy gas structure.

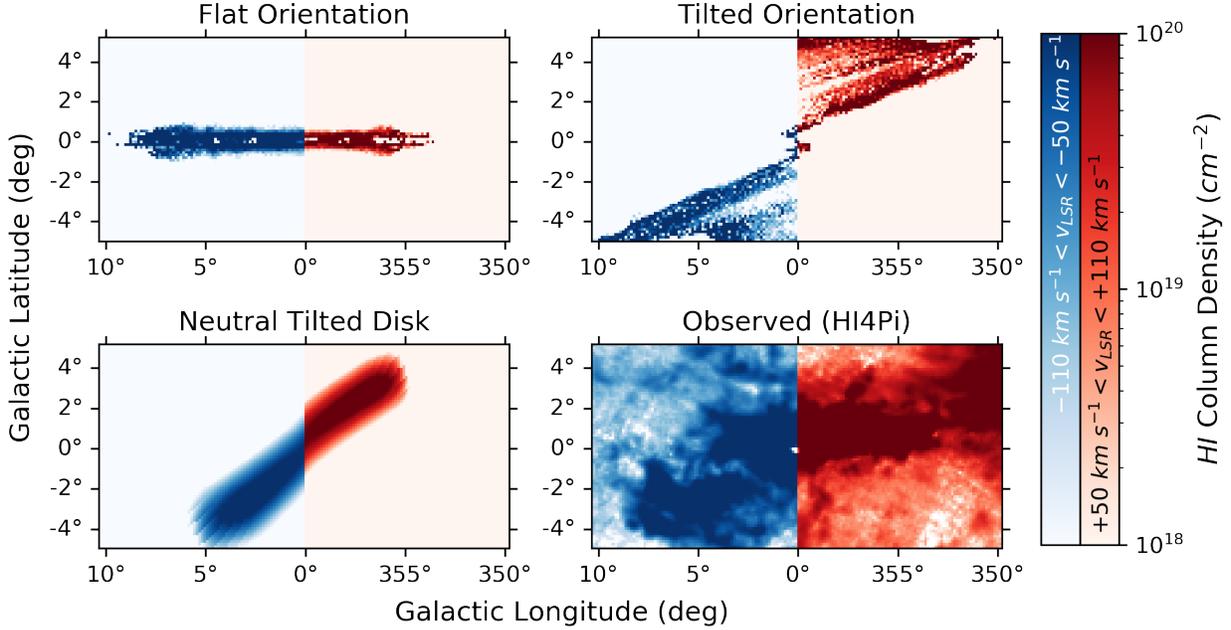

**Fig. S7. Edge-on HI Comparisons.** Edge-on maps of HI column density from hydrodynamical simulations (*26*) [top left, top right], the Tilted Disk (*23*) [bottom left], and the observed 21-cm HI distribution from HI4Pi (*22*) [bottom right]. Maps are restricted to the "forbidden" velocities as shown within the colorbars. The Tilted Disk model best predicts the distribution of the observed neutral gas.

A table of the total masses and vertical thickness for both the hydrodynamical and Tilted Disk models is given in **Table S2**. For the hydrodynamic simulations, the "total" mass of ionized and neutral gas is measured within the white ellipse shown in **Fig S3** and is compared to the total mass of the Tilted Disk model. The mass of the gas interior to the blue polygon in **Fig S3** covers a similar longitude and velocity range as our Baade's window mass estimates. The atomic gas mass (neutral plus ionized) within comparable elliptical regions for both the geometrical and hydrodynamical models is similar, approximately 15 x $10^6$ solar masses, but with much different ionization fractions: 80% ionized for the geometrical model and less than 1% ionized for the hydrodynamical model. This is likely attributable to the lack of hydrogen-ionizing photons in the radiation field used in the simulations. The atomic gas mass estimates for gas with negative velocities over the Baade's window longitude range is more discrepant, 0.8 x $10^6$ solar masses (geometrical) vs. 0.01x $10^6$ solar masses (hydrodynamical), with a similar discrepancy in the ionization fraction.

Since the gas in the hydrodynamical simulations does not have a well-defined scaleheight, we define the thickness to be the height, $z_{max}$, for which the average neutral or ionized gas density drops below $n_H=10^{-2}$ cm$^{-3}$. With the exception of the original HI

Tilted Disk model (which had a fixed $z_{max}$=0.28 kpc), this value increases with radius and is larger for the ionized gas than the neutral gas. However, the thickness of the gas distribution for the geometrical model is nearly an order of magnitude larger than what is found for the hydrodynamical simulations.

|  | Tilted Disk | Hydrodynamical Model |
|---|---|---|
| Gas Mass | ($10^6$ Solar Masses) | ($10^6$ Solar Masses) |
| $M^{tot}(H^0)$ | 3.1 ± 0.3 | ~14.7 |
| $M^{tot}(H^+)$ | 12 (+4/-3) | ~0.09 |
| $M^{BW}(H^0)$ | 0.30 ± 0.01 | ~0.01 |
| $M^{BW}(H^+)$ | 0.37 (+0.12/-0.09) | ~0.00007 |
| Vertical Extent | (kpc) | (kpc) |
| $z_{max}(H^0)$ [center] | 0.28 | 0.01 |
| $z_{max}(H^+)$ [center] | 0.95 ± 0.15 | 0.02 |
| $z_{max}(H^0)$ [max radius] | 0.28 | 0.12 |
| $z_{max}(H^+)$ [max radius] | 1.92 (+0.49/-0.43) | 0.09 |

**Table S2. Comparison of Masses and Vertical Extent of Models.** The total mass $M^{tot}$ of neutral ($H^0$) or ionized ($H^+$) gas comes from the full geometrical model [left column] or within an elliptical region shown in Fig S5 for the hydrodynamical model [right column]. The "Baade's window" mass $M^{BW}$ is the mass in a limited range of longitude and projected velocity (see text) where lower optical extinction allows for a direct comparison to the neutral and ionized components. The maximum vertical gas extent, defined by where the particle density drops below $n_H=10^{-2}$ cm$^{-3}$, is measured both near the center and maximum radius of the distribution.

To summarize, our reference hydrodynamical model produces a gas layer that is more neutral, thinner, and more planar than is observed. The addition of a radiation field with a high enough flux of hydrogen ionizing photons may obviate the first issue. Additional sources of vertical pressure support, e.g. cosmic rays and turbulence, may produce a thicker distribution. But the cause of the tilted distribution remains unresolved. It may also be necessary to consider how these features evolve over this history of the simulation since we may not be observing a steady-state configuration.

*Software Used*

This work uses the following open source python software: *astropy (54, 55)*, *numpy (56)*, *matplotlib (57)*, *seaborn (58)*, *emcee (51)*, *dustmaps (50)*, *extinction (59)*, *spectral-cube (60)*, *whampy (41)*, *modspectra ([https://modspectra.readthedocs.io](https://modspectra.readthedocs.io))*